\documentclass[sigconf,screen]{acmart}
\acmConference[EASE 2026]{The 30th International Conference on Evaluation and Assessment in Software Engineering}{Tue 9 - Fri 12 June 2026}{Glasgow, United Kingdom}

\usepackage{enumitem}

\usepackage{nicefrac}

\usepackage{todonotes}
\usepackage{multirow}

\usepackage{cleveref}

\usepackage{framed}

\definecolor{lightgray}{gray}{0.95}

\newenvironment{rqanswer}
  {%
   \MakeFramed{\advance\hsize-\width\FrameRestore}}
  {\endMakeFramed}

\newcommand{\RQone}{What characterizes review activity for AI-generated PRs?}
\newcommand{\RQtwo}{How does this activity differ from review of human-authored pull requests in the same repositories?}

\usepackage{makecell}

\usepackage[english]{babel}

\usepackage{booktabs}

\usepackage{amsmath}
\usepackage{graphicx}

\title[These Aren't the Reviews You're Looking For How Humans Review AI-Generated Pull Requests]{These Aren't the Reviews You're Looking For \\ How Humans Review AI-Generated Pull Requests}
\author{Kacper	Duma} 
\orcid{0009-0000-7184-9958}
\affiliation{%
  \institution{Nicolaus Copernicus University}
  \city{Toruń}
  \country{Poland}
}
\email{kacperd@mat.umk.pl}

\author{Patryk	Wróblewski} 
\affiliation{%
  \institution{Nicolaus Copernicus University}
  \city{Toruń}
  \country{Poland}
}
\orcid{0009-0008-8701-1824}
\email{pwro@mat.umk.pl}

\author{Jagoda	Bobińska} %
\orcid{0009-0008-1382-745X}
\affiliation{%
  \institution{Nicolaus Copernicus University}
  \city{Toruń}
  \country{Poland}
}
\email{gusiabob@mat.umk.pl}

\author{Julia	Winiarska} 
\orcid{0009-0004-2926-4189}
\affiliation{%
  \institution{Nicolaus Copernicus University}
  \city{Toruń}
  \country{Poland}
}
\email{juliawiniarska@mat.umk.pl}

\author{Piotr Przymus}
\orcid{0000-0001-9548-2388}
\affiliation{%
  \institution{Nicolaus Copernicus University}
  \city{Toruń}
  \country{Poland}
}
\email{piotr.przymus@mat.umk.pl}

\begin{document}

\begin{abstract}
We analyze code review interactions for AI-generated pull requests (PRs) on GitHub using the AIDev dataset and compare them to human-authored PRs within the same repositories. We find that most AI-generated PRs receive no review and, when reviewed, are largely dominated by AI agents rather than humans. Human-authored PRs are more likely to receive human-only review and to attract direct human feedback. In contrast, reviews of AI-generated PRs more often take the form of automation-mediated interaction, with human involvement frequently expressed through agent steering rather than standalone evaluation. These results indicate systematic differences in how review activity is structured in agentic workflows and raise challenges for interpreting review metrics as indicators of human oversight in large-scale mining studies.
\end{abstract}

\maketitle

\section{Introduction}

Code review is a central practice in software engineering, traditionally described as an evaluator-driven process in which humans inspect code, discuss alternatives, and iteratively converge toward improved solutions. Empirical work, however, shows that real-world reviews often diverge from this idealized model. Contemporary reviews tend to be lightweight and brief discussions, few reviewers, and rapid approvals
rather than in-depth technical critique~\cite{rigby2013convergent}. A recent survey further highlights that modern practice prioritizes speed, pragmatism, and asynchronous, tool-supported workflows over exhaustive evaluation~\cite{badampudiModernCodeReviews2023}. As a result, even human-authored pull requests (PRs) frequently receive limited substantive feedback.

The emergence of \emph{AI-generated pull requests} adds another dimension to this streamlined landscape. Coding agents now autonomously propose changes across open-source repositories, but little is known about how humans respond to these contributions. Prior work provides early evidence on the variability of automated review tools~\cite{sun2025aicodeactions}, yet almost nothing is known about \emph{human} review behaviour when the author is an AI agent. Do developers offer meaningful technical feedback on these changes, or do interactions primarily consist of acknowledgments, workflow comments, or procedural steering? And how does this compare to reviews of human-authored PRs within the same repositories?

This problem is important because AI-generated pull requests are becoming increasingly common, yet we don't have a clear understanding of how developers engage with and evaluate contributions authored by AI agents. Without this knowledge, it is difficult to assess the quality of collaboration and broader impact of AI on software development practices.

To investigate these questions, we conduct a large-scale empirical study of \emph{human review activity} on AI-generated (``agentic'') PRs in GitHub repositories with at least 100~stars avalible in AIDev dataset~\cite{li2025aidev}. For analysis, we distinguish review interactions that clearly fall into \emph{non-evaluative} categories---such as CI-related messages, or steering commands for AI agents---from all remaining comments, which we treat as \emph{evaluative} for the purposes of quantification (this category varies in depth but is adequate for capturing broad review differences).\\

\noindent\textbf{Research Question:}\\
\noindent\textbf{RQ1:} \RQone\\
\noindent\textbf{RQ2:} \RQtwo\\

\noindent\textbf{Contributions:}
\begin{itemize}[leftmargin=1em, itemsep=1pt, topsep=2pt]
    \item Large-scale analysis of human review activity on AI-generated and human-written PRs fromSS~ AIDev repositories ($\geq100 \star$).
    \item Classification of review interactions into non-evaluative vs. other comments to quantify human involvement.
    \item Replication package: \url{https://github.com/ncusi/reviewing-ai-generated-prs-ease2026-short}
\end{itemize}






\section{Related Work}

\subsection{AI-Assisted Code Review}
Prior research explores automated support for code review, including
defect detection, style enforcement, and maintainability improvement
\cite{bacchelliExpectationsOutcomesChallenges2013a,
sadowskiModernCodeReview2018b, badampudiModernCodeReviews2023,
davilaSystematicLiteratureReview2021}. Recent studies focus on LLM-based reviewers
integrated into pull request (PR) workflows.
Sun et al.~\cite{sun2025aicodeactions} analyze over 22{,}000 AI-generated
review comments on GitHub and show that actionable feedback containing concrete
code suggestions is significantly more likely to trigger code changes. Other studies report mixed productivity effects: while AI reviewers can reduce manual
effort, they may also increase comment volume without shortening review
latency \cite{watanabeUseChatGPTCode2024, cihan2024automatedcodereviewpractice,
adalsteinssonRethinkingCodeReview2025}. Tufano et al. \cite{tufano2024codereviewautomationstrengths} conduct a qualitative analysis of automated code review techniques and characterize the types of changes for which current approaches succeed or fail. They report that existing models handle simple modifications more reliably than changes requiring broader contextual understanding of the code base. Turzo et al. \cite{turzo2023automatedclassificationcodereview} propose a deep learning-based classifier for review comments and show that fine-grained categorization provides better informative code review analytics than coarse-grained metrics based on counts.

\subsection{Human--AI Interaction in Code Review}
Several works investigate how developers perceive and use AI-assisted review.
Qualitative studies indicate that AI feedback can reduce social friction but
introduce additional cognitive load related to validating AI suggestions
\cite{alami2025humanmachinesoftwareengineers}. Developers therefore tend to treat AI comments
as advisory rather than authoritative, selectively integrating them based on
context and trust.

\subsection{Agentic Software Engineering}
The concept of \emph{agentic software engineering} extends AI assistance toward
autonomous agents acting as participants in development workflows. Tang et
al.~\cite{tangCodeAgentAutonomousCommunicative2024} propose a multi-agent system
capable of iterative and communicative code review. Conceptual work emphasizes
the need for new evaluation criteria for such systems, including transparency
and alignment with human team norms \cite{rokemTenSimpleRules2024}. Gong et al. ~\cite{gong2026analyzingmessagecodeinconsistencyai} analyze the inconsistency in agent-authored PR and report that PRs with misaligned descriptions are less likely to be accepted and take longer to merge.

\subsection{Review Dynamics in Pull Requests}
Empirical studies of human code review analyze review latency, reviewer
participation, and comment characteristics, providing important baselines for
understanding collaborative review processes
\cite{rigby2013convergent, cetinReviewCodeReviewer2021}. However, direct
comparisons between traditional human review workflows and AI-driven agent
review processes remain limited. Pirouzkhah et al. \cite{pirouzkhah2026thevalueofeffectivepullrequestdescription} examine PR descriptions at scale and report that certain description elements, especially explicit statements of desired feedback, are associated with greater reviewer engagement and higher likelihood of merge.

\subsection{Review Comment Classification}
Prior work has proposed taxonomies to describe the content of code review comments. Those classifications typically distinguish between functional issues, maintainability concerns, documentation related remarks and discussion oriented feedback. More recent studies explore automated classification of review comments to support large scale analytics \cite{turzo2023automatedclassificationcodereview}. In contrast, our study does not aim to introduce a fine-grained semantic taxonomy of comment types. Rather than categorizing review content by technical topic, we focus on distinguishing interaction modes—agent steering, automation-related interaction, and other human-driven review. This coarser scheme aligns with our objective of analyzing how human oversight is expressed in agentic workflows, rather than evaluating the substantive technical quality of review feedback.

\section{Dataset and Methodology}

\begin{figure}
    \centering
    \includegraphics[  width=\linewidth,
  trim=0 0 0 1cm,
  clip, width=1\linewidth]{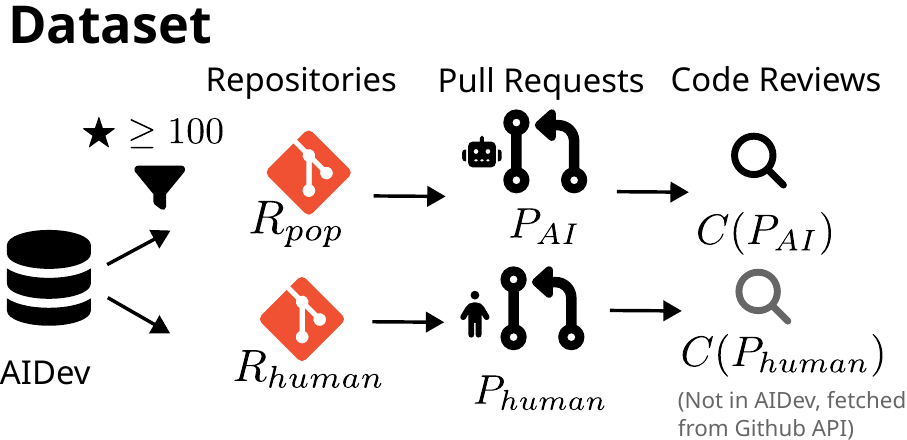}
\caption{Overview of the AIDev-based data collection pipeline. Pull requests and code reviews are mined from GitHub repositories, combining AIDev annotations with additional review data retrieved via the GitHub API.}
    \label{fig:placeholder}
\end{figure}

\begin{table*}
    \centering
\caption{PR-level overview of code review presence and reviewer composition. Review presence indicates whether a PR received any review. For reviewed PRs, reviewer composition distinguishes between agentic-only, human-only, and mixed (human + agentic) participation. R - indicates repositories subset. Percentages are relative to row totals.}
\label{tab:basicstats}
\begin{tabular}{lllrlllllll}
\toprule 
\multicolumn{3}{r}{} & \multicolumn{2}{c}{Review Presence} & \multicolumn{3}{c}{Reviewer Composition}&
\multicolumn{2}{c}{Reviews (mean $\pm$ std)}
\\
 \cmidrule(lr){4-5} \cmidrule(lr){6-8} \cmidrule(lr){9-10}
PR author & R & \# PRs & \# no review& \# with review & AI & Human & Human + AI & AI& Human\\
\midrule
AI agent & $R_{pop}$ & 33596 & 20621 (61.38\%) & 12975 (38.62\%) & 7625 (58.77\%) & 1316 (10.14\%) & 4034 (31.09\%) & 2.35 ± 1.95 & 2.19 ± 2.37 \\
AI agent & $R_{\cap}$ & 9616 & 2781 (28.92\%) & 6835 (71.08\%) & 3939 (57.63\%) & 552 (8.08\%) & 2344 (34.29\%) & 2.66 ± 2.16 & 2.19 ± 2.46 \\
Human & $R_{\cap}$ & 5574 & 1924 (34.52\%) & 3650 (65.48\%) & 1932 (52.93\%) & 920 (25.21\%) & 798 (21.86\%) & 2.05 ± 1.66 & 2.57 ± 3.32 \\
\bottomrule
\end{tabular}

\end{table*}

\label{sec:dataset}
\subsection{Dataset}

We use the AIDev dataset~\cite{li2025aidev}, which contains over 932k pull requests together with metadata identifying agentic and human authorship.
From AIDev, we derive the sets of AI-generated pull requests $P_{AI}$ and human-authored pull requests $P_{human}$, as well as the repository sets $R_{pop}$ (repositories containing at least one agent-authored PR) and $R_{human}$ (repositories containing at least one human-authored PR). We further define their intersection as
\[
R_{\cap} = R_{pop} \cap R_{human}.
\]

AIDev includes full review activity for repositories in $R_{pop}$.
For repositories in $R_{human}$ whose reviews were not covered by AIDev, we
retrieve the corresponding review and discussion records via the GitHub
REST API. 
Let $C$ denote the union of all collected review comments. We partition
$C$ into $C_{AI}$ (review comments authored by agents) and $C_{human}$
(review comments authored by humans). For any PR set $P$ we write $C(P)$
for all reviews targeting PRs in $P$, with
\[C(P) = C_{AI}(P) \cup C_{human}(P).\]
These sets define the slices used in our research questions:\\
$\textbf{RQ1}\; C(P_{AI}\mid R_{pop})$\\
$\textbf{RQ2}\; C(P_{AI}\mid R_{\cap})\ \text{vs.}\ C(P_{human}\mid R_{\cap}), R_{\cap}= R_{pop}\cap R_{human}$.

Table~\ref{tab:basicstats} summarizes the analyzed pull requests after filtering. Filtering details follow.

\subsection{Filtering}

\subsubsection{Filtering Human PRs}
The AIDev dataset provides a stratified sample of human-authored pull requests designed to match the characteristics of AI-generated pull requests. 
During manual inspection, however, we observed that some pull requests labeled as human-authored were in fact created by automated agents.
To address this issue, we applied a two-step filtering procedure to clean the set $P_{\text{human}}$.

First, we identified accounts that were mislabeled as human but in fact corresponded to bots. These accounts were initially detected based on the presence of the ``[bot]'' pattern in the username. Importantly, \emph{every account matching this pattern was manually inspected and verified} to ensure correct classification.

Second, we filtered out pull requests whose bodies contained AI-related keywords (e.g., \emph{PR-Codex}, \emph{Copilot}, \emph{Claude}, \emph{Devin}, \emph{Cursor}). All pull requests matching these patterns were subsequently manually inspected to confirm that they were in fact created by AI agents.

Overall, we removed 1{,}044 pull requests from $P_{\text{human}}$.

\subsubsection{Filtering Human Code Reviews}

Similarly, we cleaned the set of human-authored review comments to obtain a reliable $C_{\text{human}}$. 
Although AIDev explicitly labels many bot accounts, some automated review mechanisms appear under the account type ``User'' and are therefore are not identified as ``Bot''. 

To detect such cases, we applied a heuristic filtering procedure. We first restricted attention to accounts with user type = ``User''. Among these, we flagged usernames exhibiting bot-like patterns (e.g., ``[bot]'', ``ci'', ``automation''). To reduce false positives, we additionally required that such accounts have at least $k$ interactions (e.g., $k = 10$). All flagged accounts were then manually inspected to confirm their automated nature.

Repositories involving such accounts were excluded, yielding a human-only set $R_{\text{human}}$ used for comparisons with $R_{\text{pop}}$ and $R_{\cap}$. All identified accounts were labeled as ``Bot Other''.

In total, we removed 588 misclassified human-authored comments from $C(P_{AI})$ and 539 from $C(P_{\text{human}})$.


\begin{table*}[]
    \centering
    \caption{Distribution of comment-level review comments per PR (with $\geq$1 review), distinguishing agentic vs. human authorship. Human comments are categorized ($C_{\text{human}}$). $R$ denotes the repository subset. Percentages are coresponding to row totals.}
\label{tab:comment_distribution}
\begin{tabular}{lllrlllll}
\toprule
 \multicolumn{2}{r}{} & \multicolumn{3}{c}{Reviews} & \multicolumn{3}{c}{Human review categories  ($\kappa$ on $C_{\text{human}}$)} \\
 \cmidrule(lr){3-5} \cmidrule(lr){6-8}
 PR author & R & All & $C_{AI}$ & $C_{human}$ & Agentic &   Automation & Human review\\
\midrule
AI agent & $R_{pop}$ & 39122 & 28004 (71.58\%) & 11118 (28.42\%) & 3154 (28.37\%) & 789 (7.1\%) & 7175 (64.53\%) \\
AI agent & $R$ & 23081 & 16727 (72.47\%) & 6354 (27.53\%) & 1647 (25.92\%) & 543 (8.55\%) & 4164 (65.53\%) \\
Human & $R$ & 10001 & 5589 (55.88\%) & 4412 (44.12\%) & 72 (1.63\%) & 212 (4.81\%) & 4128 (93.56\%) \\
\bottomrule
\end{tabular}
\end{table*}
\section{Classification of Comments}
To help us perform our review analysis, we designed a rule-based classifier to assign each human-authored comment to one of three categories. In this paper, the classifier is used only as a helper tool. It is based on regular expressions derived from a data-driven analysis of the corpus. To construct the rules, we grouped comments by normalized text, inspected the most frequent forms, and iteratively expanded the patterns to cover common variants. Human-authored review classifier assignees each comment exactly one category: 

\[
\kappa : C_{human} \rightarrow
\{\text{agent},\ \text{automation},\ \text{human\_review}\}.
\]
\textbf{Agentic.}  
Agent-related comments correspond to human steering of LLM-based coding
assistants. These are detected by pairing agent names (e.g., \texttt{copilot},
\texttt{coderabbit}, \texttt{devin}, \texttt{claude}, \texttt{gemini},
\texttt{sourcery-ai}) with imperative verbs such as ``run'', ``fix'',
``rebase'', ``test'', or ``update''. We include both inline commands
(e.g., ``@coderabbit fix lint failure'') and comments that start with an agent
address.

\textbf{Automation.}  
Automation/CI comments are matched using a separate set of patterns targeting
chatops commands, CI-service prefixes, and known bot output formats. 
These include commands such as
``/azp run'', ``/test'', ``/rebase'', auto-formatting bots, CLA reminders,
coverage annotations, and other system-generated diagnostic notices. The
defining characteristic is interaction with infrastructure rather than with
LLM agents.

\textbf{Human Review.}
All remaining human-authored comments that do not match the agent or automation patterns are labeled as human review. In the initial design, workflow-related remarks and substantive evaluative feedback were treated as separate categories. However, empirical analysis of the corpus revealed substantial lexical and functional overlap between these two types of comments, and attempts to separate them introduced unstable boundaries without analytical benefit. We therefore merged them into a single category. This merged class includes both detailed code-review comments about functionality, design, correctness, performance, style, and maintainability, as well as short process-oriented statements that facilitate the pull-request workflow (e.g., “lgtm”, “done”, “thanks”, pings such as “@user”, merge/close notices, reminders to rerun tests or formatting, and instructions to clean or revert specific files). 
This category is treated as the default class for human-driven interaction within pull requests.

\subsection{Classification rules}
Throughout development, we repeatedly sampled false positives and false negatives to refine category boundaries and ensure consistent separation between agent, human review and automation comments. Particular attention was paid to ambiguous cases where infrastructure and agent - triggering commands, and human-authored feedback might overlap.  The complete implementation, all regular-expression patterns, and representative examples for each category are included in our replication package.

The classifier is implemented using structured regular expressions that encode lexical and structural signals observed in the corpus. Agent-related comments are detected through the co-occurrence of a closed lexicon of agent identifiers (e.g., Copilot, Claude, Gemini, Devin, CodeRabbit) and a predefined set of imperative action verbs (e.g., run, fix, rebase, test, update, review). A comment is classified as agent-directed if it matches either an address–command structure (e.g., “@agent fix lint”) or a command–entity co-occurrence pattern within the same textual span.

Automation and CI-related comments are identified using prefix-based and format-based expressions capturing chatops commands (e.g., “/azp run”, “/rebase”), infrastructure triggers, and standardized bot output markers such as coverage annotations, conflict warnings, or CLA reminders. These patterns rely on initial markers and service specific tokens specific for  infrastructure interaction.

All regular expressions are evaluated deterministically and in fixed order. Agent patterns are applied first, followed by automation patterns. Any comment that does not match these patterns is assigned to the default human review category. This default class therefore encompasses both substantive evaluative feedback (e.g., comments on functionality, design, correctness, or style) and procedural workflow-related remarks (e.g., “lgtm”, merge notices).

\subsection{Validation of the Classifier}
To ensure reliability, we conducted a manual validation of the regular-expression rules on the datasets used for both RQs. For each dataset, we drew stratified random samples with equal numbers of instances from each predicted category of human comments. In total, 800 comments were manually inspected and re-labeled to assess classification correctness. After merging workflow and review into a single human review category, the aggregated three-class confusion matrix (agentic, bot, human review) is as follows: 
\begin{table}[h]
\centering
\caption{Regex classifier confusion matrix.}
\label{tab:confusion_matrix}
\begin{tabular}{lccc}
\toprule
 & Agentic & Automation & Human \\
\midrule
Agentic & 189 & 0   & 10 \\
Automation & 0   & 200 & 7  \\
Human   & 11  & 0   & 383 \\
\bottomrule
\end{tabular}
\end{table}

Overall accuracy equals \textbf{96.5\%} (772/800).
Given the large sample size, residual classification errors are unlikely to  affect the observed aggregate proportions.


Per-class performance is consistently high: \textbf{Agentic} (precision 94.5\%, recall 95.0\%), \textbf{Automation} (precision 100\%, recall 96.6\%), and \textbf{Human Review} (precision 95.8\%, recall 97.2\%).

Misclassifications occur mainly between agentic and human review, reflecting lexical overlap between imperative forms and informal discourse. Confusion with bot is negligible.

These indicate that the regex rules are consistent and that the three-class scheme preserves strong discriminative performance with a simpler structure.

\section{Results}
In this study, we first analyze AI-generated pull requests from repositories in $R_{pop}$ as from AIDev dataset~\cite{li2025aidev}.
We then compare code review practices between agent-authored and human-authored pull requests originating from the same set of repositories ($R_{\cap}$; see Section~\ref{sec:dataset} for details).

Table~\ref{tab:basicstats} summarizes pull-request--level review presence and reviewer composition, while Table~\ref{tab:comment_distribution} reports comment-level review activity and the distribution of human review categories.

Across all settings, review activity is generally lightweight, with reviewed pull requests receiving on average 2--3 comments with standard deviations below 3.5.

\subsection{Agentic Pull Requests in $R_{pop}$}

We first examine review activity for agent-authored pull requests in popular repositories ($R_{pop}$). In total, $33{,}596$ agent-authored PRs are included in this subset. At the pull-request level (see~\Cref{tab:basicstats}), 61.38\% (20{,}621) receive no recorded review activity, while 38.62\% (12{,}975) receive at least one review.
Among those agent-authored PRs that are reviewed, 58.77\% (7{,}625) are reviewed exclusively by agents, 10.14\% (1{,}316) receive human-only review, and 40.34\% (4{,}034) involve mixed human--agent participation. 

Viewed from the perspective of observable human involvement, 84.0\% (\nicefrac{28246}{33596}) of agent-authored PRs either receive no recorded review or are reviewed exclusively by agents, whereas 15.9\% (\nicefrac{5350}{33596}) exhibit some form of human participation. We note that the absence of recorded review activity does not imply the absence of human oversight; maintainers may inspect pull requests without leaving traceable comments. Our results capture observable review interaction as reflected in the PR history.

At the comment level (see~\Cref{tab:comment_distribution}), review discussions are dominated by agent-authored comments. Of 39{,}122 total review comments, 71.58\% (28{,}004) are authored by agents and 28.42\% (11{,}118) by humans. Within human comments, 64.53\% (7{,}175) are categorized as direct human review (evaluative and workflow-related feedback), 28.37\% (3{,}154) correspond to agent-steering interactions, and 7.10\% (789) are automation-related messages.

\subsection{Agentic and Human Pull Requests in $R_{\cap}$}

We compare agent-authored and human-authored pull requests within the same repositories ($R_{\cap}$) using Pearson’s chi-square tests on categorical count data.
To account for multiple comparisons, we apply the Benjamini--Hochberg false discovery rate correction and adopt a conservative significance threshold of $p < 0.001$.
 We report the chi-square statistic together with its degrees of freedom (number of independent categories in the contingency table) and effect sizes using Cramér’s $V$.

At the pull-request level (see~\Cref{tab:basicstats}), agent-authored pull requests are reviewed slightly more often than human-authored pull requests (71.08\% vs.\ 65.48\%).
This difference is statistically significant but small in magnitude ($\chi^2(1)=51.44$, $p<10^{-3}$, $V=0.06$).

Viewed from the perspective of observable human involvement, 69.9\% (\nicefrac{6720}{9616}) of agent-authored PRs in $R_{\cap}$ either receive no recorded review or are reviewed exclusively by agents, whereas 30.1\% (\nicefrac{2896}{9616}) exhibit some form of human participation.
For human-authored PRs in the same repositories, 69.2\% (\nicefrac{3856}{5574}) either receive no recorded review or are reviewed exclusively by agents, while 30.8\% (\nicefrac{1718}{5574}) involve observable human participation.
As before, the absence of recorded review activity does not imply the absence of human oversight; maintainers may inspect pull requests without leaving traceable comments.

Reviewer composition differs much more strongly between author types.
For agent-authored pull requests, most reviewed cases are agent-only (57.63\%), with mixed Human+AI reviews also common (34.29\%), and human-only reviews relatively rare (8.08\%).
For human-authored pull requests, agent-only reviews remain frequent (52.93\%), but mixed reviews are less common (21.86\%), and human-only reviews are substantially more frequent (25.21\%).
This difference is statistically significant with a moderate effect size ($\chi^2(2)=629.4$, $p<10^{-3}$, $V=0.25$).

At the comment level (see~\Cref{tab:comment_distribution}), reviews of agent-authored pull requests are dominated by agentic comments (72.47\%), while human-authored pull requests attract a substantially larger share of human comments (44.12\%).
This difference is statistically significant with a moderate effect size ($\chi^2(1)=873.7$, $p<10^{-3}$, $V=0.16$).

The strongest divergence concerns the distribution of human review comment categories.
For agent-authored pull requests, comments unrelated to steering neither agents nor bots account for 65.53\% of human review comments, compared to 93.56\% for human-authored pull requests.
Conversely, agent-steering commands are far more common when reviewing agent-authored pull requests (25.92\%) than human-authored ones (1.63\%).
This difference is statistically significant with a large effect size ($\chi^2(3)=1280$, $p<10^{-3}$, $V=0.34$).
Results remain significant under the adopted threshold.

\section{Discussion}

\subsection{RQ1: \RQone}

AI-generated pull requests exhibit limited and automation-heavy review activity. In $R_{pop}$, 61.38\% of agent-authored PRs receive no recorded review activity. Overall, 22.6\% (\nicefrac{7625}{33596}) are reviewed exclusively by agents and 15.9\% ( \nicefrac{1316+4034}{33596}) involve observable human participation. Thus, most agent-authored PRs either receive no recorded review or are evaluated exclusively through automated participation.

At the comment level, review discussions are similarly dominated by agents: 71.58\% of review comments are authored by agents. Human comments account for 28.42\% and consist primarily of direct human review (64.53\%), alongside agent-steering (28.37\%) and automation-related interaction (7.10\%). We initially distinguished workflow comments (e.g., “LGTM”) from other types of reviews, but due to frequent annotator disagreement, we merged this into direct human review.

\begin{rqanswer}
\noindent\textbf{Answer to RQ1}
AI-generated pull requests are frequently unreviewed and, when reviewed, are predominantly handled by agents or through mixed human--agent participation. Observable human involvement is limited and includes both direct review feedback and agent-steering interaction.
\end{rqanswer}

\subsection{RQ2: \RQtwo}

Within the same repositories ($R_{\cap}$), agent-authored PRs are reviewed slightly more often than human-authored PRs, although this difference is small ($V=0.06$). Importantly, the overall rate of observable human participation is nearly identical: 30.12\% for agent-authored PRs and 30.83\% for human-authored PRs.

The primary differences concern reviewer composition and interaction structure rather than the mere presence of human involvement. Human-authored PRs are substantially more likely to receive human-only review (25.21\%) than agent-authored PRs (8.08\%), whereas mixed human--agent participation is more common for agent-authored PRs (34.29\% vs.\ 21.86\%) ($V=0.25$).

At the comment level (ie. $C_{AI}$ vs $C_{Human}$), human-authored PRs attract a larger share of human comments (44.12\% vs.\ 27.53\%) ($V=0.16$). 

The strongest divergence concerns human comment categories ($V=0.34$): for human-authored PRs, 93.56\% of human comments constitute direct human review and agent-steering is nearly absent (1.63\%). In contrast, reviews of agent-authored PRs include a substantially larger share of agent-steering (25.92\%) and a lower proportion of direct human review (65.53\%).

\begin{rqanswer}
\noindent\textbf{Answer to RQ2}
Compared to human-authored PRs in the same repositories, AI-generated PRs exhibit similar overall levels of observable human participation but differ markedly in how review effort is structured, with fewer human-only reviews and substantially more agent-steering interaction.
\end{rqanswer}

\paragraph{Interpretation and Consequences.}
The strongest differences are structural rather than quantitative. While human participation occurs at comparable overall rates within the same repositories, its form changes when the author is an AI agent. Reviews of agent-authored PRs more frequently involve mixed participation and steering-oriented interaction, whereas human-authored PRs are predominantly evaluated through direct human review.

Compared to traditional human-centric workflows, configurations with no recorded human review activity provide limited evidence of explicit human evaluation in the PR history. Although maintainers may inspect pull requests without leaving comments, the absence of documented interaction reduces the traceability of review decisions and complicates the interpretation of review metrics as indicators of oversight. Observable review activity reflects interaction patterns rather than documented reasoning.

Taken together, these findings suggest that AI authorship is associated with changes in the observable structure of review workflows, even when overall levels of human participation remain similar. Our results describe recorded interaction patterns and should not be interpreted as definitive evidence of reduced human oversight.

\section{Threats to Validity}

\textbf{Construct Validity.}
Agent-authored pull requests are identified using the AIDev dataset, with additional filtering and manual verification to remove automation accounts from the human-authored set. Mislabeling may still occur. Review comments are classified using deterministic rules; short or ambiguous comments may be misclassified. Such errors may affect exact proportions but are unlikely to change overall trends.

The absence of review comments does not imply that the code was not reviewed (e.g., it may have received a silent approval). Nevertheless, we classify all PRs without comments as not reviewed, as there is no empirical basis to distinguish between these cases.

\textbf{Internal Validity.}
Differences in review activity may be influenced by factors other than authorship, such as pull request size, complexity, or repository-specific workflows. These factors are not controlled for in this study.

\textbf{External Validity.}
Our analysis is limited to GitHub repositories in the AIDev dataset and, for RQ2, to repositories containing both agent-authored and human-authored pull requests. Results may not generalize to other settings.

\textbf{Conclusion Validity.}
We report descriptive results and do not assess code quality or review effectiveness. Findings should be interpreted as differences in review activity.

\section{Conclusion}

This study investigates how AI-generated pull requests are reviewed in practice using large-scale AIDev dataset~\cite{li2025aidev}, comparing review activity for agent-authored and human-authored PRs within the same repositories.

AI-generated pull requests are frequently unreviewed and, when reviewed, are predominantly evaluated by agents or through mixed human--agent configurations. 
Note, that within the same repositories, overall levels of observable human participation are similar for agent-authored and human-authored PRs. 

At the comment level, reviews of agent-authored PRs are dominated by agent-generated comments, while human participation includes both direct review feedback and a substantial share of steering-oriented interaction. In contrast, human-authored PRs are far more likely to receive human-only review and to attract predominantly direct human feedback, whereas AI-generated PRs more often involve mixed participation and significantly higher levels of agent-steering interaction.



These findings indicate that AI-generated contributions are associated not with the disappearance of human involvement, but with measurable changes in how review effort is expressed and documented. In agentic workflows, observable review activity does not necessarily correspond to direct human evaluation, and silent or undocumented reviews limit the traceability of human oversight. While using PR comments for agent steering can reuse existing infrastructure and shorten feedback loops, it also blurs the boundary between evaluation and interaction. As a result, conventional review metrics should be interpreted with care when used as indicators of human oversight and, where possible, should distinguish between human review, automation, and agent-steering activity.

Future work should examine the implications of the distinct review configurations identified here. Fully automated review loops, delegated steering-based interaction, and augmentative AI participation in human-authored PRs represent different modes of review that may vary in effectiveness and oversight characteristics. Integrating interaction-pattern analysis with measures of review quality and post-merge outcomes could help clarify these effects and inform the development of appropriate metrics for assessing human oversight in increasingly automated development environments.

\bibliographystyle{plain}
\bibliography{paper}

\end{document}